\pdfoutput=1
\documentclass[11pt]{article}

\usepackage[]{emnlp2021}

\usepackage{times}
\usepackage{latexsym}
\usepackage{microtype}
\usepackage{graphicx}
\usepackage{booktabs}  
\usepackage{multirow}
\usepackage{subfigure}
\usepackage{verbatim}
\usepackage{CJKutf8} 
\usepackage{amssymb}
\usepackage{amsmath}

\usepackage{algorithm}
\usepackage{algorithmic}
\usepackage{makecell}
\usepackage{flushend}
\usepackage[T1]{fontenc}
\usepackage[utf8]{inputenc}

\usepackage{microtype}

\title{DebiasGAN: Eliminating Position Bias in News Recommendation \\with Adversarial Learning}
\author{Chuhan Wu$^\dagger$~~~~Fangzhao Wu$^\ddagger$~~~~\textbf{Yongfeng Huang}$^\dagger$\\
    $^\dagger$Department of Electronic Engineering \& BNRist, Tsinghua University, Beijing 100084, China  \\
     $^\ddagger$Microsoft Research Asia, Beijing 100080, China\\
  \tt{\{wuchuhan15,wufangzhao\}@gmail.com, yfhuang@tsinghua.edu.cn}
  }

\begin{document}
\maketitle
\begin{abstract}

News recommendation is important for improving news reading experience of users.
Users' news click behaviors are widely used for inferring user interests and predicting future clicks.
However, click behaviors are heavily affected by the biases brought by the positions of news displayed on the webpage.
It is important to eliminate the effect of position biases on the recommendation model to  accurately target user interests.
In this paper, we propose a news recommendation method named \textit{DebiasGAN} that can effectively eliminate the effect of position biases via adversarial learning.
We use a bias-aware click model to capture the influence of position bias on click behaviors, and we use a bias-invariant click model with random candidate news positions to estimate the ideally unbiased click scores.
We apply adversarial learning techniques to the hidden representations learned by the two models to help the bias-invariant click model capture the bias-independent interest of users on news.
Experimental results on two real-world datasets show that \textit{DebiasGAN} can effectively improve the accuracy of news recommendation by eliminating position biases.

\end{abstract}

\section{Introduction}

Accurate news recommendation is critical for improving users' online news reading experience~\cite{wu2020mind}.
Existing news recommendation methods mainly use users' news click behaviors to model their interest and predict future clicks~\cite{wang2018dkn,wu2019,wu2019nrms,ge2020graph,hu2020graph,wu2020user}.
For example, \citet{okura2017embedding} used GRU to infer user interests from clicked news, and compute click scores based on the relevance between candidate news and user interests.
\citet{wu2019nrms} used multi-head self-attention  to model user interests based on clicked news, and predict click scores by matching candidate news and user interests.
However, news click behaviors are heavily influenced by position biases, i.e., the  positions of news when displayed on a webpage~\cite{chen2020bias}.
News displayed at top positions are more likely to be clicked than those displayed inconspicuously~\cite{baeza2018bias}.
Thus, models directly learned on biased click data may be inaccurate in user interest modeling and click prediction~\cite{yi2021debiasedrec}.

\begin{figure}[!t]
  \centering 
      \includegraphics[width=0.9\linewidth]{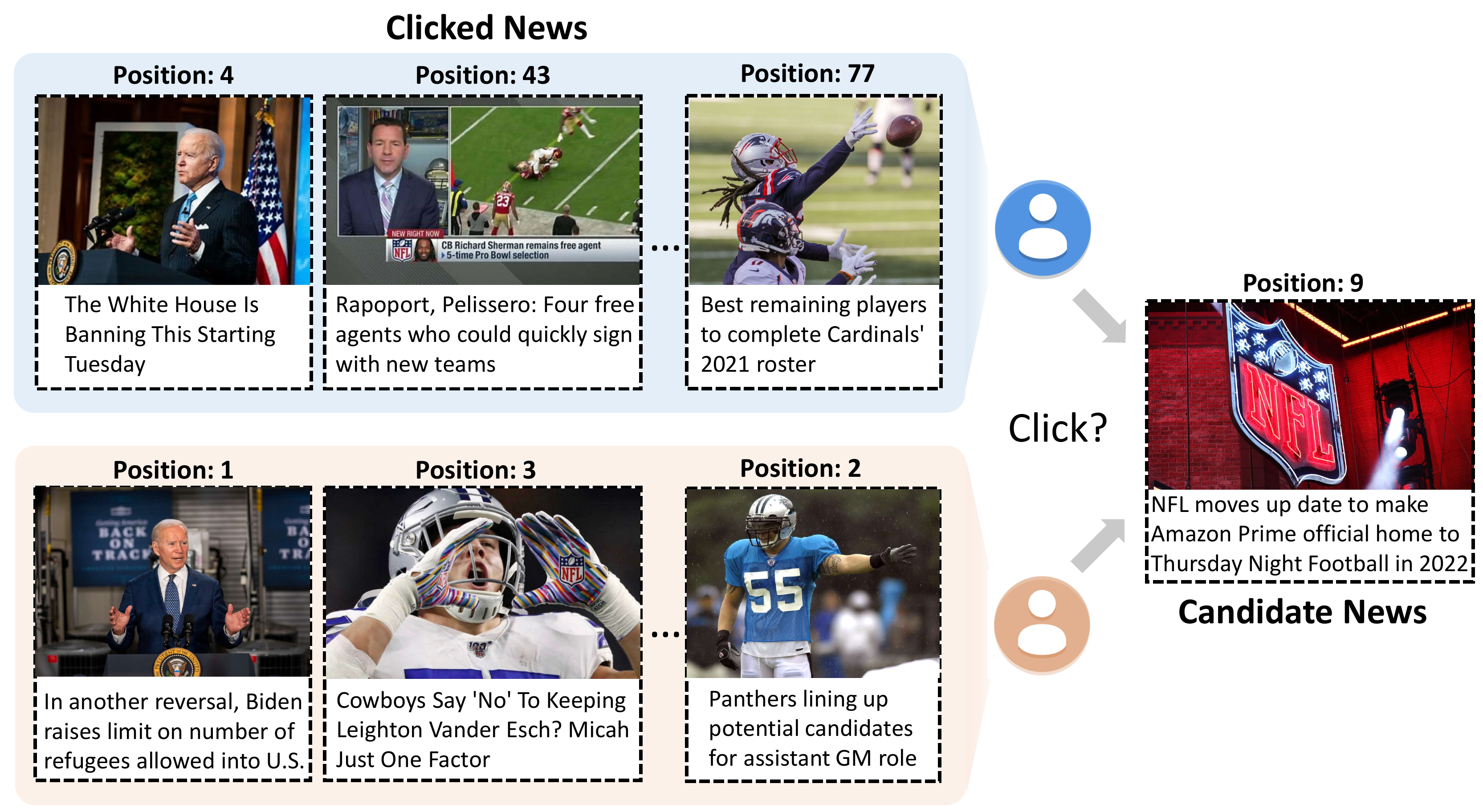}
      \vspace{-0.05in}
  \caption{The clicked news of two users and a candidate news displayed to them.}\label{fig.exp} \vspace{-0.1in}
\end{figure}

There are many works on eliminating position biases in recommendation~\cite{wang2018position,guo2019pal,sato2020unbiased,wu2021unbiased}. For example, \citet{wang2018position} proposed a regression-based EM method to estimate the bias propensity to help learn unbiased models.
\citet{guo2019pal} proposed a bias-aware model learning method that decomposes the overall click probability into a user-item relevance score  and the probability of an item being seen by a user.
In these methods position bias is modeled in a user-independent way.
However, a news displayed at the same positions may generate different impacts on different users.
For example, as shown in Fig.~\ref{fig.exp}, the first user clicks news displayed at both high and low positions, while the second user mainly clicks news displayed at top positions.
Thus, the candidate news displayed at the same medium position is more likely to be clicked by the first user rather than the second user.
Thus, incorporating user preferences into bias modeling can help better model and eliminate the effects of position biases. 

In this paper, we propose a news recommendation method named \textit{DebiasGAN}, which can effectively reduce the effect of position biases on news recommendation via adversarial learning.
In our approach, we use a bias-aware click  model to capture the effects of position biases on click behaviors, and we use a bias-invariant click model with randomized positions of candidate news to estimate the ideally unbiased click scores.
Both models can consider the interactions between candidate news and clicked news to incorporate user preferences into bias modeling.
In addition, we apply adversarial learning techniques to the hidden representations learned by the two models to encourage the bias-invariant click model to capture position bias-independent interests of users on candidate news. 
Experiments on two real-world datasets show that \textit{DebiasGAN} can effectively reduce position biases to improve  the  accuracy  of  news  recommendation.

\section{DebiasGAN}\label{sec:Model}

\begin{figure}[!t]
  \centering 
      \includegraphics[width=0.94\linewidth]{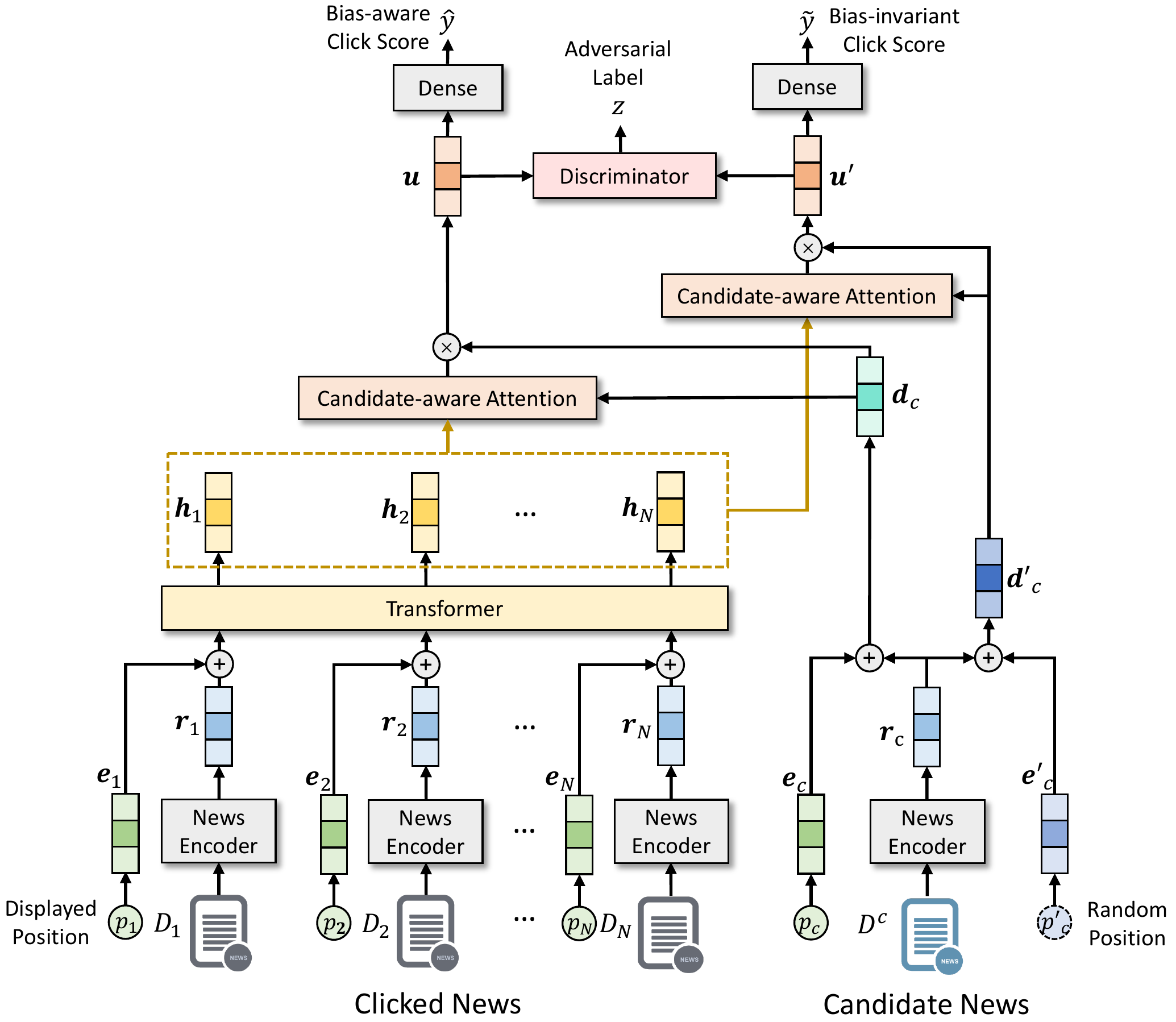}
 \vspace{-0.05in}
  \caption{Model architecture of \textit{DebiasGAN}.}\label{fig.model} \vspace{-0.1in}
\end{figure}

Next, we introduce our proposed \textit{DebiasGAN} approach for news recommendation.
Its architecture is shown in Fig.~\ref{fig.model}.
It contains a bias-aware click model that captures the position bias effect on click behaviors and a bias-invariant click model to model bias-invariant user interest on candidate news.
Their details are introduced as follows.

\subsection{Bias-aware Click Model}

The bias-aware click model aims to compute bias-aware click scores by modeling the effects of position biases on click behaviors.
We denote the $N$ historical clicked news of a user as $[D_1, D_2, ..., D_N]$.
The candidate news is denoted as $D^c$.
We use a news encoder to learn semantic representations of news from their texts.
Motivated by~\cite{wu2019nrms}, we use the Transformer~\cite{vaswani2017attention} model as the news encoder.
We denote the hidden representation of the clicked news $D_i$ and candidate news $D^c$ as $\mathbf{r}_i$ and $\mathbf{r}_c$, respectively.
To model the impact of position biases on user interest modeling, we incorporate the embedding of the displayed positions of clicked news.
We denote the position of the clicked news $D_i$ as $p_i$ (starts from 1).
In order to reduce the sparsity of positions, we quantize each position $p_i$ by $\hat{p}_i=\lceil\sqrt{p_i-1} \rceil$.
We convert the quantized position $\hat{p}_i$ into its embedding $\mathbf{e}_i$, and add it to the semantic news embedding $\mathbf{r}_i$ to obtain bias-aware news representations.
We also add position embedding of the candidate news (denoted as $\mathbf{e}_c$) to  $\mathbf{r}_c$ to obtain a bias-aware candidate news representation $\mathbf{d}_c$.
The bias-aware representations of clicked news are further process by a behavior Transformer, which  captures the relations between click behaviors to help better model interests.
We denote its output as $\mathbf{H}=[\mathbf{h}_1, \mathbf{h}_2, ..., \mathbf{h}_N]$.

Since the same position may generate different impacts on different users, we incorporate user preferences into the modeling of the effects of position bias on click behaviors.
More specifically, we use a candidate-aware attention network to select important click behaviors to accurately model user interests on the candidate news displayed at a given position.
We denote the bias-aware user interest representation with respect to the candidate news as $\mathbf{u}$, which is formulated as follows:
\begin{equation}\small
    \mathbf{u}=\mathbf{d}_c \odot [\text{softmax}(\mathbf{d}_c^\top \tanh(\mathbf{W}_c\mathbf{H}+\mathbf{w}_c))\times \mathbf{H}],  
\end{equation}
where $\mathbf{W}_c$ and $\mathbf{w}_c$ are parameters, and $\odot$ means element-wise product.
We further predict a bias-aware click score $\hat{y}$ based on $\mathbf{u}$, which is formulated as $\hat{y}=\mathbf{w}^\top\mathbf{u}$, where $\mathbf{w}$ is a parameter vector.
This score indicates the predicted probability of a user clicking on a candidate news given the position of candidate news and the user's personal preference on the content and position of this news.

\subsection{Bias-invariant Click Model}

The bias-invariant click model aims to estimate the ideally unbiased click scores.
It shares the news encoder and the behavior  Transformer with the bias-aware click model, while it adds the embedding of a random candidate news position (denoted as $\mathbf{e}'_c$) to the candidate news representation $\mathbf{r}_c$ to form a bias-randomized representation $\mathbf{d}'_c$.\footnote{In  test phase, we add a fixed default position embedding.}
We use another candidate-aware attention network that uses $\mathbf{d}'_c$ as attention query to learn bias-invariant user interest representation $\mathbf{u}'$ on the candidate news as follows:
\begin{equation}\small
    \mathbf{u}'=\mathbf{d}'_c \odot [\text{softmax}({\mathbf{d}'_c }^\top \tanh(\mathbf{W}'_c\mathbf{H}+\mathbf{w}'_c))\times \mathbf{H}],  
\end{equation}
where $\mathbf{W}'_c$ and $\mathbf{w}'_c$ are parameters.
Since the candidate news position is randomly generated, the bias-invariant click model is encouraged to model user interests that are independent of displayed positions of news. 
We predict the bias-invariant click score  by $\tilde{y}=\mathbf{w}^\top \mathbf{u}'$, where the parameter $\mathbf{w}$ is shared with the bias-aware click model.

\subsection{Debiasing with Adversarial Learning}

To further help the bias-invariant click model estimate the unbiased click scores, we propose to apply adversarial learning techniques to the bias-aware and bias-invariant user interest representations.
We use a discriminator to classify whether $\mathbf{u}$ or $\mathbf{u}'$ is learned by the bias-aware click model.
The adversarial label is predicted by the discriminator as:
$z=\sigma(\mathbf{w}_p^T\mathbf{u}-\mathbf{w}_n^T\mathbf{u}'),$
where $\mathbf{w}_p$ and $\mathbf{w}_n$ are parameters, and $\sigma$ is the sigmoid function.
The adversarial loss function $\mathcal{L}_A$ we used is written as $\mathcal{L}_A=-\log(z)$.
By propagating the negative gradients of the adversarial loss, both bias-aware and bias-invariant click models are encouraged to learn similar representations of user interests on candidate news.
Thus, the distributions of their predicted click scores are expected to be similar, and thereby the effects of position biases on click prediction can be effectively mitigated.

\subsection{Model Training}

Finally, we introduce the training of  \textit{DebiasGAN}.
Following~\cite{wu2019npa,wu2020mind}, we construct training samples via negative sampling and we use crossentropy as the loss function for click prediction~\cite{wu2019nrms}.
We denote the click prediction losses of the bias-aware and bias-invariant click models as $\mathcal{L}_B$ and $\mathcal{L}_D$, respectively.
The unified loss $\mathcal{L}$ to be optimized is formulated as:
\begin{equation}\small
    \mathcal{L}=\mathcal{L}_B+\mathcal{L}_D-\alpha \mathcal{L}_A,
\end{equation}
where $\alpha$ is a hyperparameter that controls the intensity of adversarial training for debiasing.

\section{Experiments}\label{sec:Experiments}

\subsection{Datasets and Experimental Settings}
We conducted experiments on two datasets collected from different news services on a commercial news platform.\footnote{Anonymized for double-blind review.}
The first dataset (denoted as \textit{News}) contains click logs collected from Oct. 12, 2019 to Nov. 9, 2019.
The logs in the last week are used for test and the rest are for training and validation (logs in the last 3 days).
The second dataset (denoted as \textit{Uniform}) includes click logs collected from Jul. 12, 2020 to Feb. 1, 2021.
The training and validation sets are constructed from biased click logs, while the test set is composed of user click behaviors in the last month on uniformly and randomly ranked news lists.
The statistics of \textit{News} and \textit{Uniform} are shown in Table~\ref{table.dataset}.\footnote{We present some analysis on positions in supplements.}

\begin{table}[h]
\centering
\resizebox{0.48\textwidth}{!}{
\begin{tabular}{cccccc}
\Xhline{1.5pt}
               & \# News & \# Users & \# Impressions & \# Click & \#  Pos.  \\ \hline
\textit{News}  & 97,646  & 310,469  & 580,000    & 917,839  & 1,173            \\
\textit{Uniform} & 1,277,315 & 479,281   & 529,482        & 808,365  & 667             \\ 
\Xhline{1.5pt}
\end{tabular}
}
\vspace{-0.05in}
	\caption{Statistics of the \textit{News} and \textit{Uniform} datasets.}\label{table.dataset}\vspace{-0.1in}

\end{table}

Following many prior works~\cite{wu2019npa,wu2019nrms,wang2020fine} we use Glove~\cite{pennington2014glove} embeddings in the news encoder. 
The model optimizer is  Adam~\cite{kingma2014adam}. 
The coefficient $\alpha$ is 0.5.
The hyperparameters are tuned on validation sets.\footnote{Full hyperparameter settings are in supplements.}
We use AUC, MRR, nDCG@5 and nDCG@10 to evaluate model performance.
We repeat each experiment 5 times.
The average scores with standard deviations are reported.

\begin{table*}[h]
\centering

\resizebox{0.92\linewidth}{!}{
\begin{tabular}{c|cccc|cccc}
\Xhline{1.5pt}
\hline
\multirow{2}{*}{\textbf{Methods}} & \multicolumn{4}{c|}{\textbf{News}}                                                                                                                                        & \multicolumn{4}{c}{\textbf{Uniform}}                                                                                                                                     \\ \cline{2-9} 
                                  & AUC                                      & MRR                                      & nDCG@5                                   & nDCG@10                                  & AUC                                      & MRR                                      & nDCG@5                                   & nDCG@10                                  \\ \hline
EBNR        & 61.46$\pm$0.16                              & 20.44$\pm$0.16                              & 20.32$\pm$0.22          & 25.34$\pm$0.23          & 60.39$\pm$0.20                              & 27.05$\pm$0.19                              & 36.54$\pm$0.22          & 42.02$\pm$0.24          \\ 
DKN        & 61.25$\pm$0.20                              & 20.18$\pm$0.19                              & 20.14$\pm$0.23          & 25.03$\pm$0.26          & 60.14$\pm$0.21                              & 26.89$\pm$0.17                              & 36.39$\pm$0.19          & 41.88$\pm$0.22          \\ 
NAML       & 62.59$\pm$0.14                              & 21.40$\pm$0.13                              & 21.61$\pm$0.16          & 26.56$\pm$0.18   &   61.27$\pm$0.17                              & 28.36$\pm$0.16                              & 37.71$\pm$0.19          & 43.26$\pm$0.20          \\ 
NPA       & 62.67$\pm$0.19                              & 21.48$\pm$0.18                              & 21.69$\pm$0.20          & 26.63$\pm$0.22          & 61.30$\pm$0.20                              & 28.38$\pm$0.18                              & 37.75$\pm$0.17          & 43.27$\pm$0.19          \\ 
NRMS       & 62.87$\pm$0.12                              & 21.64$\pm$0.11                              & 21.85$\pm$0.13          & 26.79$\pm$0.14          & 61.52$\pm$0.16                              & 28.65$\pm$0.14                              & 37.92$\pm$0.15          & 43.46$\pm$0.17          \\ 
FIM       & 63.02$\pm$0.13                              & 21.73$\pm$0.12                              & 21.93$\pm$0.14          & 26.90$\pm$0.15          & 61.67$\pm$0.14                              & 28.74$\pm$0.15                              & 38.03$\pm$0.16          & 43.58$\pm$0.18          \\ \hline
IPW        & 63.04$\pm$0.15                              & 21.74$\pm$0.15                              & 21.96$\pm$0.16          & 26.98$\pm$0.18          & 61.96$\pm$0.14          & 29.11$\pm$0.13          & 38.34$\pm$0.16          & 43.80$\pm$0.17          \\
REM        & 63.16$\pm$0.18                              & 21.81$\pm$0.15                              & 22.15$\pm$0.18          & 27.20$\pm$0.19          & 62.29$\pm$0.16          & 29.34$\pm$0.15          & 38.59$\pm$0.17          & 44.03$\pm$0.19          \\
PAL        & 63.22$\pm$0.14          & 21.85$\pm$0.13          & 22.21$\pm$0.12          & 27.25$\pm$0.16          & 62.34$\pm$0.16          & 29.41$\pm$0.15          & 38.66$\pm$0.14          & 44.10$\pm$0.17          \\ \hline
DebiasGAN*      & \textbf{63.94}$\pm$0.14 & \textbf{22.58}$\pm$0.16 & \textbf{22.97}$\pm$0.15 & \textbf{28.02}$\pm$0.13 & \textbf{63.40}$\pm$0.14 & \textbf{30.52}$\pm$0.12 & \textbf{39.80}$\pm$0.16 & \textbf{45.19}$\pm$0.15 \\ \hline
 \Xhline{1.5pt}
\end{tabular}
}
 \vspace{-0.05in}
\caption{Results of different methods. *T-test  shows the improvements are significant ($p<0.05$).} \vspace{-0.1in}
\label{result}
\end{table*}

\subsection{Performance Comparison}

We compare \textit{DebiasGAN} with many  baselines.
The baseline methods for news recommendation include:
(1) \textit{EBNR}~\cite{okura2017embedding}, an embedding-based news recommendation method; (2) \textit{DKN}~\cite{wang2018dkn},  deep knowledge network for news recommendation; (3) \textit{NAML}~\cite{wu2019}, attentive multi-view learning for news recommendation; (4) \textit{NPA}~\cite{wu2019npa},  news  recommendation with personalized attention; (5) \textit{LSTUR}~\cite{an2019neural},  news recommendation with long- and short-term user representations; (6) \textit{NRMS}~\cite{wu2019nrms}, multi-head self-attention for news recommendation; (7) \textit{FIM}~\cite{wang2020fine},  fine-grained interest matching for news recommendation.
 We also compare several  methods for eliminating position biases, including (1) \textit{IPW}~\cite{joachims2017unbiased,wu2021unbiased}, using inverse propensity weighting when learning model on biased click logs; (2) \textit{Reg-EM}~\cite{wang2018position}, a regression based EM method to estimate the propensity weight; (3) \textit{PAL}~\cite{guo2019pal}, a position bias-aware learning method for CTR prediction.
 For fair comparison, in these methods we use the same bias-aware click model with our approach.
 The results are shown in Table~\ref{result}.
 We find that the models with debiasing techniques outperform those directly learned from biased click logs.
 This is because reducing the effect of position biases can help  learn more accurate click prediction model.
 In addition, our approach outperforms other debiasing methods, and the improvement on the \textit{Uniform} dataset is greater.
 This is because our approach can consider the user preference on position biases to better model and eliminate position biases.
 Moreover, our approach uses adversarial learning to help model unbiased user interests.
 Thus, our model achieves better performance than baseline methods especially when the click data is unbiased.
 
 \begin{figure}[!t]
	\centering 
	\includegraphics[width=0.68\linewidth]{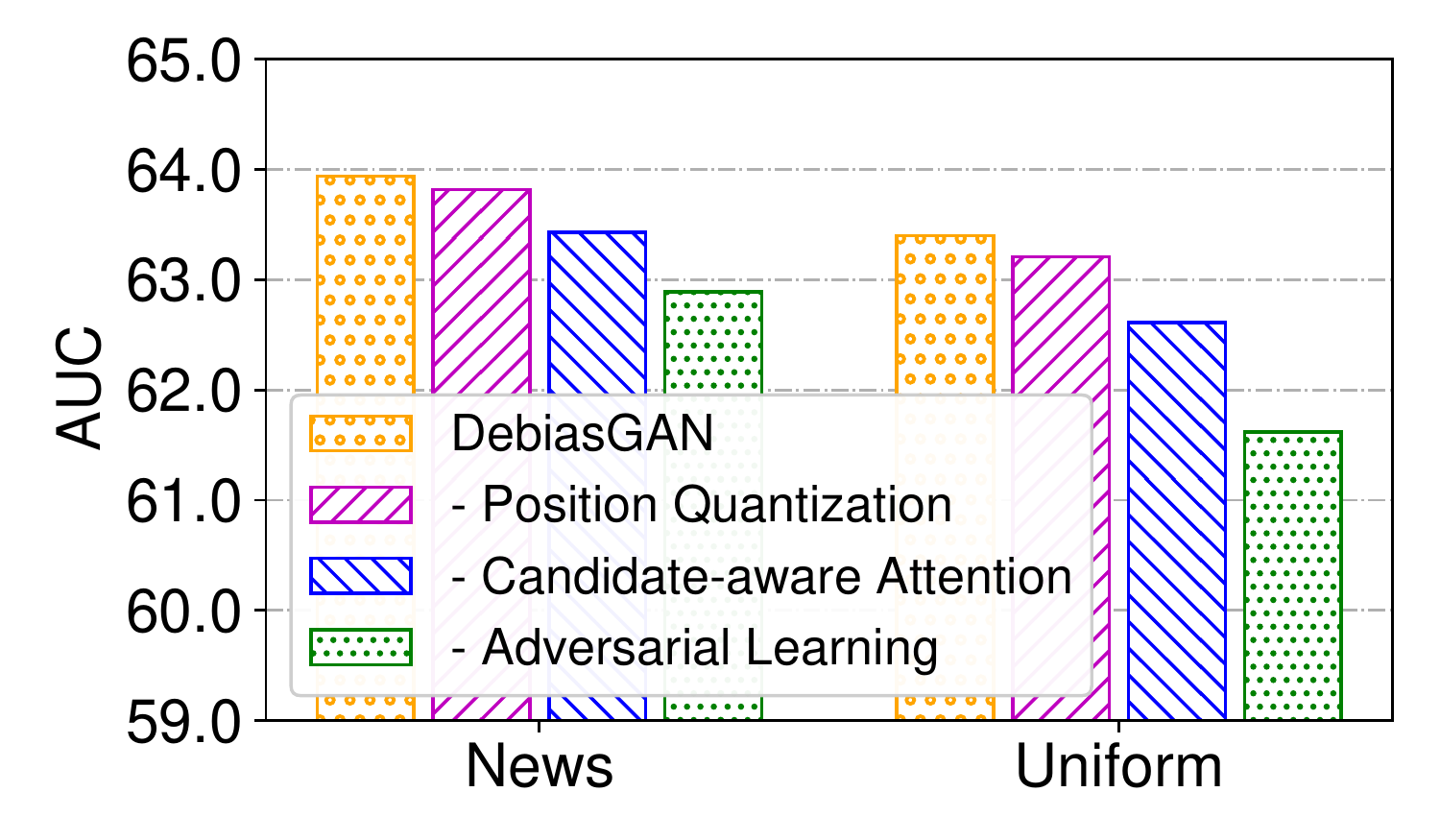} \vspace{-0.05in}
\caption{Effect of core components in \textit{DebiasGAN}.}\label{fig.d1}\vspace{-0.1in}
\end{figure}

\subsection{Ablation Study}

We verify the effectiveness of several core techniques in our approach, including candidate-aware attention networks, adversarial learning, and position quantization.
The results of \textit{DebiasGAN} and its variants without one of these components are shown in Fig.~\ref{fig.d1}.
We find that both candidate-aware attention  and adversarial learning can improve the model performance, especially on the \textit{Uniform} dataset.
This may be because candidate-aware attention can help model the user interest in candidate news, and adversarial learning can help eliminate the effect of position biases on click prediction.
In addition, quantizing the position can improve the performance.
This may be because the bias effects of adjacent positions are usually similar, and quantizing positions can also reduce their sparsity to learn accurate position embeddings.

\begin{figure}[!t]
	\centering 
	\includegraphics[width=0.68\linewidth]{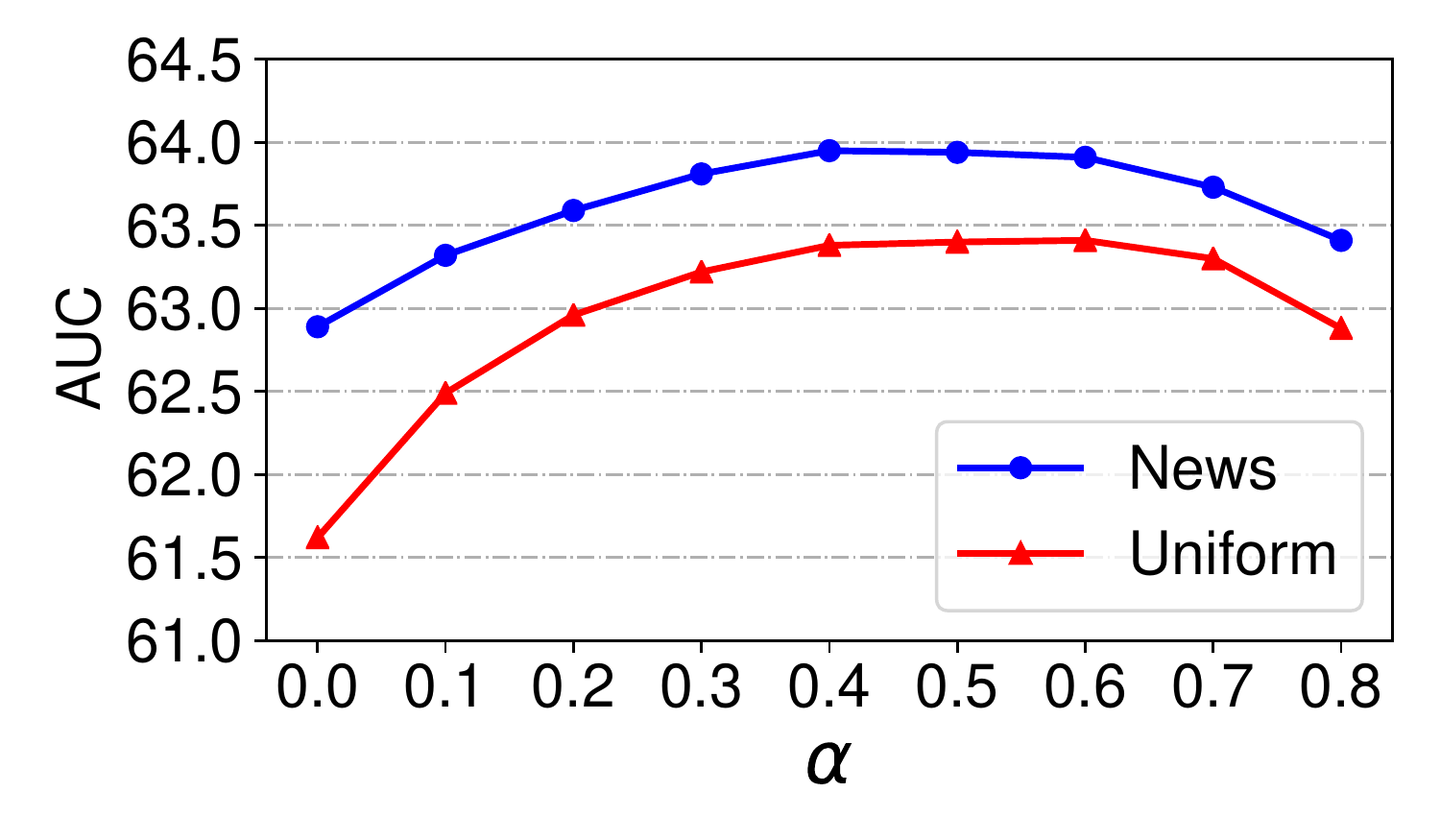} \vspace{-0.05in}
\caption{Influence of the loss coefficient $\alpha$.}\label{fig.d2}\vspace{-0.1in}
\end{figure}

\subsection{Hyperparameter Analysis}

We study the impact of the loss coefficient $\alpha$ on the model performance.
The performance in terms of AUC w.r.t. different $\alpha$ is shown in Fig.~\ref{fig.d2}.
We find that  as $\alpha$ increases, the performance first increases and then decreases.
This may be because position biases cannot be effectively removed without a sufficient intensity of adversarial gradients, while the click model may not receive adequate supervision from the main recommendation task if $\alpha$ is too large.
Thus, we choose $\alpha=0.5$ that yields good performance on both datasets.
\section{Conclusion}\label{sec:Conclusion}

In this paper, we propose a news recommendation method named \textit{DebiasGAN} that can eliminate position biases via  adversarial learning.
We propose to use a bias-aware click model to capture the effects of position biases on click behaviors and a bias-invariant click model to estimate unbiased click scores.
In addition, we apply adversarial learning techniques to the hidden user interest representations in the two models to help infer unbiased user interests on candidate news. 
Experiments on two real-world datasets validate that \textit{DebiasGAN} can effectively improve   news  recommendation performance by reducing position biases.

\bibliography{main}

\begin{thebibliography}{22}
\expandafter\ifx\csname natexlab\endcsname\relax\def\natexlab#1{#1}\fi

\bibitem[{An et~al.(2019)An, Wu, Wu, Zhang, Liu, and Xie}]{an2019neural}
Mingxiao An, Fangzhao Wu, Chuhan Wu, Kun Zhang, Zheng Liu, and Xing Xie. 2019.
\newblock Neural news recommendation with long-and short-term user
  representations.
\newblock In \emph{ACL}, pages 336--345.

\bibitem[{Baeza-Yates(2018)}]{baeza2018bias}
Ricardo Baeza-Yates. 2018.
\newblock Bias on the web.
\newblock \emph{Communications of the ACM}, 61(6):54--61.

\bibitem[{Chen et~al.(2020)Chen, Dong, Wang, Feng, Wang, and He}]{chen2020bias}
Jiawei Chen, Hande Dong, Xiang Wang, Fuli Feng, Meng Wang, and Xiangnan He.
  2020.
\newblock Bias and debias in recommender system: A survey and future
  directions.
\newblock \emph{arXiv preprint arXiv:2010.03240}.

\bibitem[{Ge et~al.(2020)Ge, Wu, Wu, Qi, and Huang}]{ge2020graph}
Suyu Ge, Chuhan Wu, Fangzhao Wu, Tao Qi, and Yongfeng Huang. 2020.
\newblock Graph enhanced representation learning for news recommendation.
\newblock In \emph{WWW}, pages 2863--2869.

\bibitem[{Guo et~al.(2019)Guo, Yu, Liu, Tang, and Zhang}]{guo2019pal}
Huifeng Guo, Jinkai Yu, Qing Liu, Ruiming Tang, and Yuzhou Zhang. 2019.
\newblock Pal: a position-bias aware learning framework for ctr prediction in
  live recommender systems.
\newblock In \emph{Recsys}, pages 452--456.

\bibitem[{Hu et~al.(2020)Hu, Li, Shi, Yang, and Shao}]{hu2020graph}
Linmei Hu, Chen Li, Chuan Shi, Cheng Yang, and Chao Shao. 2020.
\newblock Graph neural news recommendation with long-term and short-term
  interest modeling.
\newblock \emph{Inf. Process. \& Manage.}, 57(2):102142.

\bibitem[{Joachims et~al.(2017)Joachims, Swaminathan, and
  Schnabel}]{joachims2017unbiased}
Thorsten Joachims, Adith Swaminathan, and Tobias Schnabel. 2017.
\newblock Unbiased learning-to-rank with biased feedback.
\newblock In \emph{WSDM}, pages 781--789.

\bibitem[{Kingma and Ba(2015)}]{kingma2014adam}
Diederik~P. Kingma and Jimmy Ba. 2015.
\newblock Adam: A method for stochastic optimization.
\newblock In \emph{ICLR}.

\bibitem[{Okura et~al.(2017)Okura, Tagami, Ono, and
  Tajima}]{okura2017embedding}
Shumpei Okura, Yukihiro Tagami, Shingo Ono, and Akira Tajima. 2017.
\newblock Embedding-based news recommendation for millions of users.
\newblock In \emph{KDD}, pages 1933--1942.

\bibitem[{Pennington et~al.(2014)Pennington, Socher, and
  Manning}]{pennington2014glove}
Jeffrey Pennington, Richard Socher, and Christopher Manning. 2014.
\newblock Glove: Global vectors for word representation.
\newblock In \emph{EMNLP}, pages 1532--1543.

\bibitem[{Sato et~al.(2020)Sato, Takemori, Singh, and
  Ohkuma}]{sato2020unbiased}
Masahiro Sato, Sho Takemori, Janmajay Singh, and Tomoko Ohkuma. 2020.
\newblock Unbiased learning for the causal effect of recommendation.
\newblock In \emph{Recsys}, pages 378--387.

\bibitem[{Vaswani et~al.(2017)Vaswani, Shazeer, Parmar, Uszkoreit, Jones,
  Gomez, Kaiser, and Polosukhin}]{vaswani2017attention}
Ashish Vaswani, Noam Shazeer, Niki Parmar, Jakob Uszkoreit, Llion Jones,
  Aidan~N Gomez, {\L}ukasz Kaiser, and Illia Polosukhin. 2017.
\newblock Attention is all you need.
\newblock In \emph{NIPS}, pages 5998--6008.

\bibitem[{Wang et~al.(2020)Wang, Wu, Liu, and Xie}]{wang2020fine}
Heyuan Wang, Fangzhao Wu, Zheng Liu, and Xing Xie. 2020.
\newblock Fine-grained interest matching for neural news recommendation.
\newblock In \emph{ACL}, pages 836--845.

\bibitem[{Wang et~al.(2018{\natexlab{a}})Wang, Zhang, Xie, and
  Guo}]{wang2018dkn}
Hongwei Wang, Fuzheng Zhang, Xing Xie, and Minyi Guo. 2018{\natexlab{a}}.
\newblock Dkn: Deep knowledge-aware network for news recommendation.
\newblock In \emph{WWW}, pages 1835--1844.

\bibitem[{Wang et~al.(2018{\natexlab{b}})Wang, Golbandi, Bendersky, Metzler,
  and Najork}]{wang2018position}
Xuanhui Wang, Nadav Golbandi, Michael Bendersky, Donald Metzler, and Marc
  Najork. 2018{\natexlab{b}}.
\newblock Position bias estimation for unbiased learning to rank in personal
  search.
\newblock In \emph{WSDM}, pages 610--618.

\bibitem[{Wu et~al.(2019{\natexlab{a}})Wu, Wu, An, Huang, Huang, and
  Xie}]{wu2019}
Chuhan Wu, Fangzhao Wu, Mingxiao An, Jianqiang Huang, Yongfeng Huang, and Xing
  Xie. 2019{\natexlab{a}}.
\newblock Neural news recommendation with attentive multi-view learning.
\newblock In \emph{IJCAI}, pages 3863--3869.

\bibitem[{Wu et~al.(2019{\natexlab{b}})Wu, Wu, An, Huang, Huang, and
  Xie}]{wu2019npa}
Chuhan Wu, Fangzhao Wu, Mingxiao An, Jianqiang Huang, Yongfeng Huang, and Xing
  Xie. 2019{\natexlab{b}}.
\newblock Npa: Neural news recommendation with personalized attention.
\newblock In \emph{KDD}, pages 2576--2584.

\bibitem[{Wu et~al.(2019{\natexlab{c}})Wu, Wu, Ge, Qi, Huang, and
  Xie}]{wu2019nrms}
Chuhan Wu, Fangzhao Wu, Suyu Ge, Tao Qi, Yongfeng Huang, and Xing Xie.
  2019{\natexlab{c}}.
\newblock Neural news recommendation with multi-head self-attention.
\newblock In \emph{EMNLP-IJCNLP}, pages 6390--6395.

\bibitem[{Wu et~al.(2020{\natexlab{a}})Wu, Wu, Qi, and Huang}]{wu2020user}
Chuhan Wu, Fangzhao Wu, Tao Qi, and Yongfeng Huang. 2020{\natexlab{a}}.
\newblock User modeling with click preference and reading satisfaction for news
  recommendation.
\newblock In \emph{IJCAI}, pages 3023--3029.

\bibitem[{Wu et~al.(2020{\natexlab{b}})Wu, Qiao, Chen, Wu, Qi, Lian, Liu, Xie,
  Gao, Wu et~al.}]{wu2020mind}
Fangzhao Wu, Ying Qiao, Jiun-Hung Chen, Chuhan Wu, Tao Qi, Jianxun Lian,
  Danyang Liu, Xing Xie, Jianfeng Gao, Winnie Wu, et~al. 2020{\natexlab{b}}.
\newblock Mind: A large-scale dataset for news recommendation.
\newblock In \emph{ACL}, pages 3597--3606.

\bibitem[{Wu et~al.(2021)Wu, Chen, Zhao, He, Yin, and Chang}]{wu2021unbiased}
Xinwei Wu, Hechang Chen, Jiashu Zhao, Li~He, Dawei Yin, and Yi~Chang. 2021.
\newblock Unbiased learning to rank in feeds recommendation.
\newblock In \emph{WSDM}, pages 490--498.

\bibitem[{Yi et~al.(2021)Yi, Wu, Wu, Li, Sun, and Xie}]{yi2021debiasedrec}
Jingwei Yi, Fangzhao Wu, Chuhan Wu, Qifei Li, Guangzhong Sun, and Xing Xie.
  2021.
\newblock Debiasedrec: Bias-aware user modeling and click prediction for
  personalized news recommendation.
\newblock \emph{arXiv preprint arXiv:2104.07360}.

\end{thebibliography}
\bibliographystyle{acl_natbib}

%\clearpage
%\input{data/supplement}

\end{document}